\documentclass[reprint, prl, aps, twocolumn, floatfix, showkeys]{revtex4-1}

\usepackage{amssymb}
\usepackage{amsmath}
\usepackage{graphicx}
\usepackage{bm}
\usepackage{dcolumn}
\usepackage{color}
\usepackage{ulem}
\usepackage{float}
\usepackage{textcomp}
\restylefloat{table}
\usepackage{verbatim}
\usepackage{epstopdf}
\usepackage[colorlinks]{hyperref}
\usepackage[mediumspace,mediumqspace,squaren]{SIunits}

\begin{document}

\title{Observation of spontaneous valley polarization of itinerant electrons}
\date{\today}

\author{Md.\ S. Hossain}
\author{M. K.\ Ma}
\author{K.\ A. Villegas-Rosales}
\author{Y. J.\ Chung}
\author{L. N.\ Pfeiffer} 
\author{K. W.\ West}
\author{K. W.\ Baldwin}
\author{M.\ Shayegan}
\affiliation{Department of Electrical Engineering, Princeton University, Princeton, New Jersey 08544, USA}


\begin{abstract}

Memory or transistor devices based on electron's spin rather than its charge degree of freedom offer certain distinct advantages and comprise a cornerstone of spintronics \cite{Zutic.RMP.2004}. Recent years have witnessed the emergence of a new field, valleytronics, which seeks to exploit electron's valley index rather than its spin. An important component in this quest would be the ability to control the valley index in a convenient fashion. Here we show that the valley polarization can be switched from zero to one by a small reduction in density, simply tuned by a gate bias, in a two-dimensional electron system. This phenomenon arises fundamentally as a result of electron-electron interaction in an itinerant, dilute electron system. Essentially, the kinetic energy favors an equal distribution of electrons over the available valleys, whereas the interaction between electrons prefers single-valley occupancy below a critical density. The gate-bias-tuned transition we observe is accompanied by a sudden, two-fold change in sample resistance, making the phenomenon of interest for potential valleytronic transistor device applications. Our observation constitutes a quintessential demonstration of valleytronics in a very simple experiment.

\end{abstract}

\maketitle 

Electrons in several important solids, such as Si, Ge, AlAs, Bi, and single-layer materials (e.g., graphene, MoS$_2$, and WSe$_2$), possess a $valley$ degree of freedom. The electrons in these solids occupy multiple conduction-band minima, or valleys, in the  Brillouin zone. In pioneering work on two-dimensional electron systems (2DESs) in AlAs, Gunawan \textit{et al.} \cite{Gunawan.PRL.2006, Gunawan.PRB.2006, Shayegan.AlAs.Review.2006} demonstrated a direct analogy between the spin and valley degrees of freedom, and showed that the valley occupancy can be manipulated to make a functional ``valley-filter" device.  They also suggested that the valley degree of freedom might find use as a qubit for quantum computation. Rycerz \textit{et al.} \cite{Rycerz.2007}  extended this idea in 2007 to graphene and coined the term ``valleytronics". Since 2007, there has been an astonishing surge of theoretical and experimental activity on valleytronics in many material systems; for a recent review, see Ref. \cite{Schaibley.2016}. Here we report the observation of a fundamental property of an interacting, two-valley 2DES: electrons in a two-valley system suddenly all transfer to one of the valleys as the electron density is lowered below a critical value ($n_V$) (Fig. 1\textbf{a}) by tuning a voltage bias to the back gate of the sample. Such electrical control over the valley polarization opens up a new avenue in valleytronics.

Our platform is a modulation-doped AlAs quantum well, which contains a 2DES with exceptionally high quality and a valley degree of freedom. The AlAs is grown on a GaAs substrate using molecular beam epitaxy. Figures 1\textbf{b}-\textbf{e} capture some highlights of our sample and experimental setup; more details are presented in Section I of Supplementary Information (SI) \cite{SM}. As illustrated in Fig. 1\textbf{b}, in bulk AlAs six conduction-band valleys, with their minima centered at the edges of the Brillouin zone, are occupied. This leads to the occupancy of three conduction band valleys (inside the first Brillouin zone) with ellipsoidal Fermi surfaces; we refer to these valleys as X, Y, and Z (Fig. 1\textbf{b}). In our 21-nm-wide quantum well, thanks to a biaxial, compressive strain in the sample plane (caused by the slightly larger lattice constant of AlAs compared to GaAs), 2D electrons occupy only valleys X and Y, with their major axes lying in the plane along the [100] and [010] direction, respectively (Fig. 1\textbf{c}) \cite{Shayegan.AlAs.Review.2006, Chung.PRM.2018}. The electrons in each valley possess an anisotropic Fermi sea with longitudinal and transverse effective masses of $m_l = 1.1$ and $m_t = 0.20$, leading to an in-plane effective mass of $(m_t m_l)^{1/2}=0.46$. In the absence of any external strain that breaks the in-plane symmetry, the X and Y valleys are degenerate in energy. We control the valley occupancy by applying an in-plane, uniaxial strain $ \varepsilon=\varepsilon_{[100]}-\varepsilon_{[010]}$, where $\varepsilon_{[100]}$ and $\varepsilon_{[010]}$ are the strain values along [100] and [010] \cite{Shayegan.AlAs.Review.2006}; our experimental setup is shown in Fig. 1\textbf{d}. We glue the sample to a piezoelectric actuator \cite{Shayegan.APL.2003}, which deforms when a voltage ($V_{P}$) is applied and hence strains the sample glued to it. To tune the electron density ($n$), we apply voltage bias to a Ti/Au layer which is deposited on the back of the sample and serves as gate electrode.

Figure 1\textbf{e}, which shows the sample's piezoresistance at $n = 13.6 \times 10^{10}$ cm$^{-2}$ as a function of $V_{P}$, demonstrates how we tune and monitor the valley occupancy. At $V_P = V_B$, i.e. at $\varepsilon=0$ (point \textbf{B}), the 2DES exhibits isotropic transport, namely, the resistances measured along [100] and [010] ($R_{[100]}$ and $R_{[010]}$) are equal, even though the individual valleys are anisotropic. We refer to this point where the two valleys are degenerate as the ``balance" point. Note that a finite $V_P$ is often required to attain $\varepsilon=0$ because of a cooldown-dependent residual strain \cite{Shayegan.AlAs.Review.2006}. For $\varepsilon>0$, as electrons transfer from X to Y, $R_{[100]}$ decreases (black trace in Fig. 1\textbf{e}) because the electrons in Y have a small effective mass and therefore higher mobility along [100]. (Note that the total 2DES density remains fixed as strain is applied.) The resistance eventually saturates (e.g., point \textbf{C}), when all electrons are in the Y valley \cite{Shayegan.AlAs.Review.2006, Gokmen.Natphy.2010}. For $\varepsilon<0$, $R_{[100]}$ increases and saturates (e.g., point \textbf{A}) as electrons are transferred to X which has a large mass and a low mobility along [100]. As expected, the behavior of $R_{[010]}$ is opposite to that of $R_{[100]}$; see the red trace in Fig. 1\textbf{e}. Note that the resistance anisotropy ratio when only the X or Y valley is occupied is $\simeq 4$, somewhat smaller than the ratio ($\simeq 5$) of effective masses along the major and minor axes of a given valley; this is consistent with previous measurements \cite{Shayegan.AlAs.Review.2006}.

Figure 2 summarizes the highlight of our study. We monitor $R_{[100]}$ and $R_{[010]}$ at $\varepsilon=0$ as we tune the electron density $n$ by applying a voltage bias to the back gate in our sample (Fig. 2\textbf{a}). At high densities, $R_{[100]}$ and $R_{[010]}$ are equal, consistent with the valley degeneracy at $\varepsilon=0$ and the piezoresistance traces in Fig. 1\textbf{e}. As we reduce $n$, both $R_{[100]}$ and $R_{[010]}$ increase, consistent with the expectation that a reduction in density and mobility leads to higher resistance. However, when we lower the density to $n\simeq 6.3\times 10^{10}$ cm$^{-2}$, transport suddenly becomes anisotropic with $R_{[100]}$ dropping and $R_{[010]}$ rising (Fig. 2\textbf{a}). This behavior is consistent with a lifting of the valley degeneracy and the electrons abruptly transferring from the X valley to the Y valley. Note that $R_{[010]}$ becomes larger than $R_{[100]}$ by about a factor of $4$, indicating a Y-valley occupancy (see Fig. 1\textbf{e}). This behavior suggests a spontaneous valley polarization. As $n$ is further reduced, both $R_{[100]}$ and $R_{[010]}$ increase consistent with a reduction in density and mobility, but the 2DES remains anisotropic, implying that the electrons remain in the Y valley.

Figures 2\textbf{b}-\textbf{d} illustrate the temperature dependence of the valley transition. As we raise the temperature, the splitting between $R_{[100]}$ and $R_{[010]}$ for $n\lesssim6.3\times 10^{10}$ cm$^{-2}$ shrinks (Fig. 2\textbf{b}), likely because of thermal broadening of the Fermi function and the thermally-induced electron transfer between the two valleys. Notably, above $ T \simeq 1.2$ K, the 2DES becomes isotropic throughout the entire density range, leaving no sign of the valley transition. Figure 2\textbf{c} captures this evolution, mapping the resistance anisotropy, defined as $R_{[010]}/R_{[100]}$, as a function of density and temperature. It is clear in Fig. 2\textbf{c} that the anisotropy seen when $n\lesssim6.3\times 10^{10}$ cm$^{-2}$, which emerges thanks to the spontaneous valley polarization, vanishes at $T \gtrsim 1.2$ K. This disappearance can also be seen in Fig. 2\textbf{d} data taken at $n \simeq 6.0\times 10^{10}$ cm$^{-2}$, where we observe a resistance minimum in $R_{[100]}$ at low temperatures. 

If the transition described in Fig. 2 indeed stems from a spontaneous splitting of the valley degeneracy, it should be absent in a system which is, at high densities, intentionally fully valley polarized. To verify this, we apply large $|\varepsilon|$ to our sample and make the 2DES fully valley polarized; see points \textbf{A} and \textbf{C} in Fig. 1\textbf{e}. We then perform the same experiment as in Fig. 2\textbf{a}, namely monitor $R_{[100]}$ and $R_{[010]}$ as a function of density when only the X (point \textbf{A} in Fig. 1\textbf{e}) or Y (point \textbf{C} in Fig. 1\textbf{e}) valley is occupied. Figures 3\textbf{a}, \textbf{b} show such data. Both $R_{[100]}$ and $R_{[010]}$ in Figs. 3\textbf{a}, \textbf{b} exhibit monotonic rises as the density is lowered. When only the X valley is occupied, $R_{[100]}$ is larger than $R_{[010]}$ throughout the entire density range (Fig. 3\textbf{a}), and vice versa when only the Y valley is occupied (Fig. 3\textbf{b}). Importantly, the data of Figs. 3\textbf{a}, \textbf{b} do not show any signs of a density-driven transition. This is strikingly different from the abrupt transition at $n\simeq 6.3\times 10^{10}$ cm$^{-2}$ seen in Fig. 2\textbf{a}, where the valleys are degenerate at high densities.

As detailed in Section IV of SI \cite{SM}, we also studied another sample from a different wafer. The AlAs quantum well in the other sample has a slightly smaller width (20 nm instead of 21 nm). This sample also shows data (see Fig. S6) similar to Fig. 2\textbf{a}, but the critical density for the valley transition is slightly larger ($\simeq 6.9\times 10^{10}$ cm$^{-2}$ instead of $\simeq 6.3\times 10^{10}$ cm$^{-2}$). The data provide clear evidence for the reproducibility of the phenomenon we report. An intriguing aspect of our results is that, once the valley transition occurs, the electrons move to the Y valley in both samples. Understanding why the Y valley is favored requires future investigations.

Our observation of spontaneous valley polarization highlights an intriguing connection between electron-electron interaction and magnetism. It is possible that it relates to the fundamental concept of itinerant spin ferromagnetism predicted long ago by Felix Bloch \cite{Bloch.1929, Stoner.1947, AshcroftMermin, Attaccalite.PRL.2002}. When the density of an interacting electron system is sufficiently lowered so that the gain in exchange interaction energy due to the alignment of all spins outweighs the increase in the kinetic (Fermi) energy, a spontaneous transition to full spin polarization should occur. Inter-electron interaction is traditionally quantified using a dimensionless parameter $r_s$, defined as the average distance between electrons in units of the effective Bohr radius; equivalently, $r_s$ is also a measure of the ratio of the Coulomb to Fermi energies. Larger $r_s$ values are indicative of strong electron-electron interaction. For a 2DES, $r_s= (m e^2/4 \pi \hbar^2 \kappa \varepsilon_0)/ (\pi n)^{1/2} \propto m/\kappa $, where $m$ is the electron effective mass and $\kappa$ is the dielectric constant. Our material platform AlAs has a relatively large electron band effective mass ($m=0.46$, in units of the free electron mass) and small dielectric constant ($\kappa=10$), rendering the electrons in AlAs effectively very interacting. (In comparison, 2D electrons in GaAs have $m = 0.067$ and $\kappa=13$.) Our experimental results indeed demonstrate that the valley polarization occurs when $r_s$ exceeds a large value, $r_s \simeq 20$ (at $n \simeq 6.3\times 10^{10}$ cm$^{-2}$) when the electrons are strongly interacting. 

It is important to note that, in the same AlAs 2DES, we observe another transition to a fully-spin-polarized state at yet lower densities, $n \simeq 2.0\times 10^{10}$ cm$^{-2}$ ($r_s\simeq 35$) \cite{Hossain.Spin.Bloch.2020}. This behavior corroborates the qualitative analogy between the spin and valley degrees of freedom, as demonstrated previously \cite{Gunawan.PRL.2006, Gunawan.PRB.2006, Shayegan.AlAs.Review.2006}. However, it is intriguing that the density below which the spontaneous valley transition occurs ($n \simeq 6.30\times 10^{10}$ cm$^{-2}$) is about three times larger than its spin counterpart ($n \simeq 2.0\times 10^{10}$ cm$^{-2}$) \cite{Hossain.Spin.Bloch.2020}. This finding invites future efforts to understand why interaction manifests differently for the spin and valley degrees of freedom. Another noteworthy observation is that, our 2DES exhibits a metallic temperature dependence when the valley transition occurs. As we lower the density, the temperature dependence turns insulating below a density of $n \simeq 3.2\times 10^{10}$ cm$^{-2}$, and then the 2DES becomes fully spin polarized at $n \simeq 2.0\times 10^{10}$ cm$^{-2}$ \cite{Hossain.Spin.Bloch.2020}. These observations further highlight the role of spin and valley degrees of freedom in the metal-insulator transition problem, as emphasized in previous studies \cite{Gunawan.nphys.2007}. They also allow us to provide a rich phase diagram for the different ground states of our interacting, dilute 2DES as a function of electron density; see Section III of SI \cite{SM} for more details. 

In a broader context, electron-electron interaction in 2DESs is known to lead to ferromagnetism at high perpendicular magnetic fields when electrons occupy highly-degenerate, discrete, Landau energy levels. This so-called quantum Hall ferromagnetism is seen both for spin and valleys, when the Lande $g$-factor or the valley splitting is tuned through zero, or when two Landau levels with different spin and/or valley degree of freedom are made to cross \cite{Sondhi.PRB.1993, Maude.PRL.1996, DePoortere.Science.1999, Shkolnikov.PRL.2002, Lai.PRL.2004, Shkolnikov.PRL.2005, Padmanabhan.PRL.2010}. Very recently, a spontaneous spin polarization, akin to the Bloch ferromagnetism, was reported for composite fermion quasi-particles near a half-filled Landau level in GaAs \cite{Hossain.preprint}. Interaction-induced ferromagnetism at zero magnetic field has also been of tremendous interest recently, particularly in strongly-correlated electrons in narrow-bandwidth systems such as twisted bilayer graphene \cite{Sharpe.science.2019}, graphene-hBN heterostructures \cite{Serlin.science.2019}, and bilayer transition metal dichalcogenides \cite{Tang.nature.2020, Regan.nature.2020}. The origin of ferromagnetism in these systems is complex and may require an interplay between the Coulomb interaction and the moir\'e potential \cite{Zhang.preprint.2020}. Our observation of spontaneous valley polarization, in contrast, is made in a rather simple 2DES. While, as discussed above, Bloch/Stoner ferromagnetism may account for our observation, it is possible that the anisotropy of the two conduction-band valleys plays a crucial role. For example, we might be observing a transition to an electronic Ising nematic phase in which the four-fold rotational symmetry is broken spontaneously \cite{Abanin.PRB.2010, Fradkin.ARCMP.2010}.


Our results presented here demonstrate the electrical control of valley polarization and tuning of valley index via a small change in electron density. Moreover, the polarization transition leads to a substantial (about a factor of two) change in sample resistance (Fig. 2\textbf{a}). Combined, these phenomena illustrate a new valleytronics device, namely a simple transistor whose operation is based on the gate-bias manipulation of the valley polarization.


\section*{Data Availability Statement}

Data that support the plots within this paper and other findings of this study are available from the corresponding author upon reasonable request.

\section*{Acknowledgments}
We acknowledge support through the U.S. Department of Energy Basic Energy Science (Grant No. DEFG02-00-ER45841) for measurements, and the National Science Foundation (Grants No. DMR 1709076), No. ECCS 1906253, and No. MRSEC DMR 1420541), and the Gordon and Betty Moore Foundation’s EPiQS Initiative (Grant No. GBMF9615 to L. N. P.) for sample fabrication and characterization. We also thank J. K. Jain, H. D. Drew, S. A. Kivelson, and M. A. Mueed for illuminating discussions.  

\section*{Author contributions}
M. S. H. prepared the samples, performed the low-temperature measurements, and analyzed the data. Y. J. C., L. N. P., K. W. W., and K. W. B. grew the quantum well samples via molecular beam epitaxy. M. K. M. and K. A. V. R. helped with the measurements. M. S. H. and M. S. co-wrote the manuscript with input from all co-authors.

\begin{figure*}[t!]
\includegraphics[width=1\textwidth]{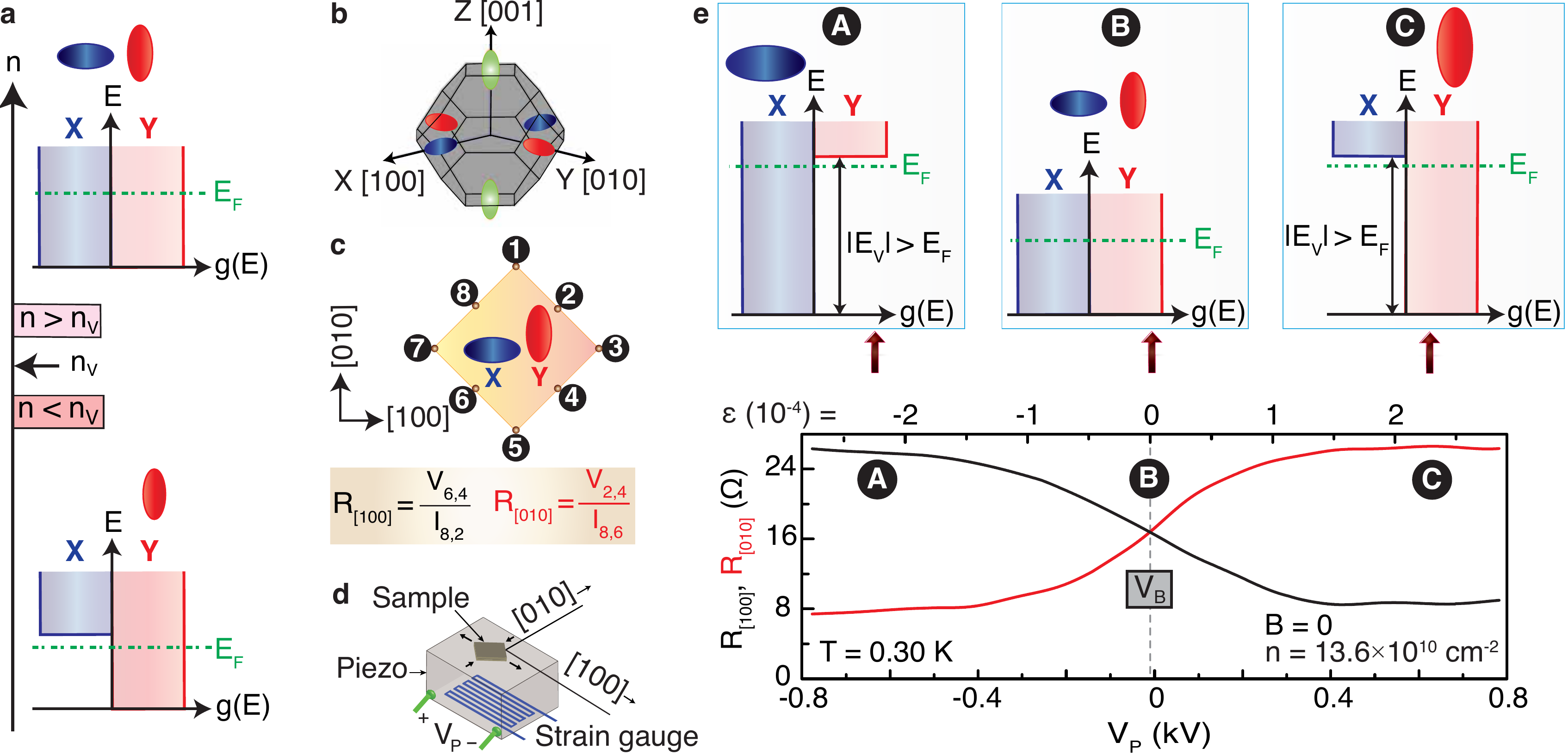}
\caption{\label{fig:Fig1} Summary of our experimental finding and material platform harboring a valley degree of freedom. \textbf{a}, Density-of-states [$g(E)$] diagrams highlighting a transition from a valley-degenerate to a valley-polarized state when the 2DES density is lowered below a critical value, $n_V$. \textbf{b}, First Brillouin zone and Fermi surfaces of the lowest-energy bands for bulk AlAs, showing the X, Y, and Z valleys; [100], [010], and [001] refer to the crystallographic directions. \textbf{c}, Projection of the Fermi seas of our AlAs 2DES where X and Y valleys are equally occupied. Electrical contacts to the sample are denoted by 1-8. Measurement configurations for $R_{[100]}$ and $R_{[010]}$ are also shown. For $R_{[100]}$, we pass current from contact 8 to 2 and measure the voltage between contacts 6 and 4. For $R_{[010]}$, the current is passed from contact 8 to 6 and we measure the voltage between contacts 2 and 4. \textbf{d}, Our experimental setup, showing an AlAs sample mounted on a piezoelectric actuator. By applying voltage bias ($V_P$) across the actuator's leads, we can tune the valley occupancy. \textbf{e}, Bottom panel: Piezoresistance of the sample at a density $n = 13.6\times 10^{10}$ cm$^{-2}$ as a function of $V_P$ and in-plane strain $\varepsilon$ (top axis). \textbf{A} - \textbf{C} mark the valley occupation. The ``balance" point, where the two (X and Y) valleys are degenerate, is achieved at $V_P = V_B = -10$ V. Top panels show density-of-states diagrams as a function of applied strain. }
\end{figure*}

\begin{figure*}[t!]
\includegraphics[width=1\textwidth]{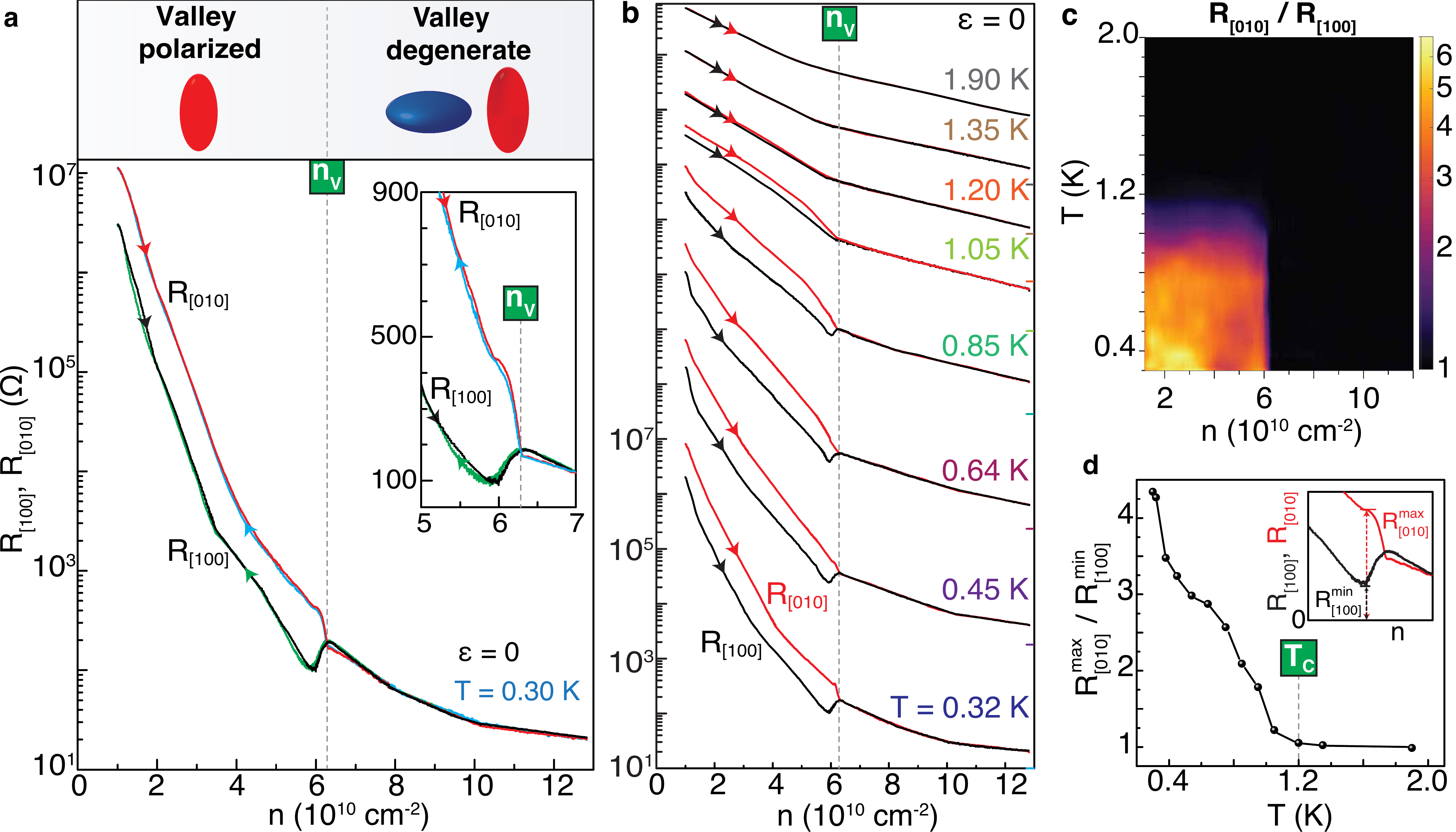}
\caption{\label{fig:Fig2} Signature of a spontaneous valley polarization in a valley-degenerate 2DES. \textbf{a}, Resistances along [100] and [010] directions plotted as a function of density. At $\varepsilon=0$, X and Y valleys are degenerate at high densities and the 2DES is isotropic. However, as the density is lowered to $n \simeq 6.3\times 10^{10}$ cm$^{-2}$, $R_{[100]}$ and $R_{[010]}$ suddenly split, signaling a valley transition. The inset zooms in near the splitting. Density sweep directions are marked with arrows. The hysteresis between the sweep directions is very small. The upper panel illustrates the spin and valley configurations as a function of density. \textbf{b}, Temperature dependence of $R_{[100]}$ and $R_{[010]}$, exhibiting the disappearance of their spontaneous splitting above $\simeq 1.2$ K. Traces are vertically offset for clarity and the temperature is marked for each set of traces. Color-coded horizontal markings on the right denote 10 $\Omega$ resistance for each temperature. \textbf{c}, Resistance anisotropy ($R_{[010]}/R_{[100]}$) as a function of density and temperature. The anisotropy sets in at $n \lesssim 6.3\times 10^{10}$ cm$^{-2}$ ($r_s \gtrsim 20$) and $T \lesssim 1.2$ K, signaling a spontaneous lifting of the valley degeneracy. The 2DES is isotropic in the rest of the ($n$, $T$) parameter space. \textbf{d}, A cross-section of the anisotropy map taken at $n \simeq 6.0\times 10^{10}$ cm$^{-2}$ where we observe a minimum in $R_{[100]}$.}
\end{figure*} 

\begin{figure}[t!]
\centering
\includegraphics[width=0.62\textwidth]{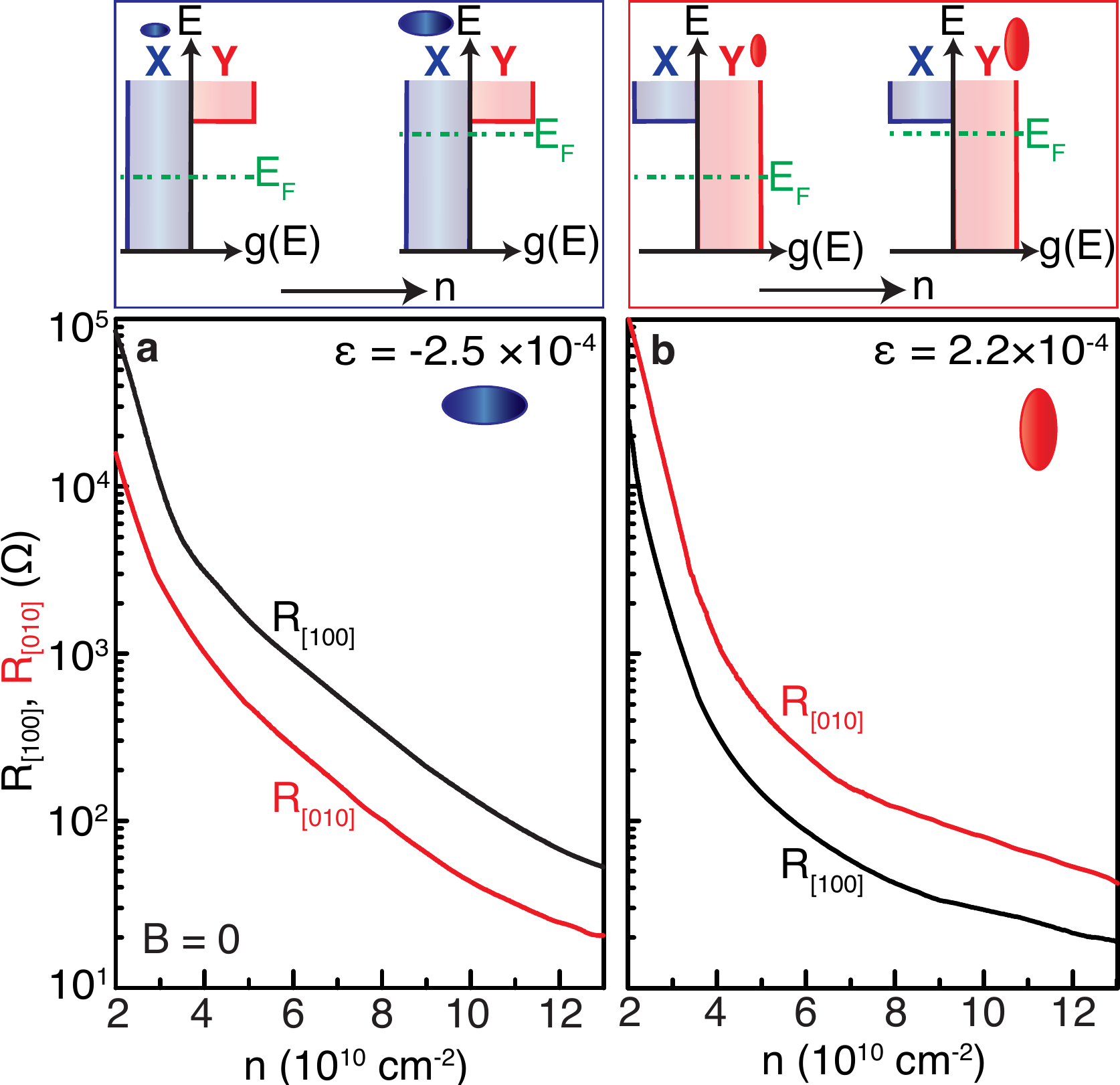}
\caption{\label{fig:Fig3} Examining the possibility of a valley transition in fully-valley-polarized 2DESs under large magnitudes of strain. Resistances along [100] and [010] directions are shown vs. electron density $n$ where only X (\textbf{a}) or only Y (\textbf{b}) valley is occupied; the values of applied strain are given in the panels. $R_{[100]}$ and $R_{[010]}$ in both cases exhibit a monotonic dependence on $n$, maintaining a similar anisotropy throughout the entire density range, and showing the absence of a valley transition. Schematics in the top panels illustrate the corresponding density-of-states diagrams as a function of $n$.
}
\end{figure} 

\end{document}